\documentclass[aps,prd,superscriptaddress,showpacs,twocolumn]{revtex4-1}
\usepackage{graphicx}
\usepackage[caption=false]{subfig}
\usepackage{epstopdf}
\usepackage{amsmath}
\usepackage{amsfonts} 
\usepackage{amssymb}
\usepackage{latexsym}
\usepackage{hyperref}
\usepackage[english]{babel}
\usepackage[utf8]{inputenc}
\usepackage[colorinlistoftodos]{todonotes}
\usepackage{xcolor}
\usepackage{slashed}
\usepackage{comment}
\usepackage{feynmp}
\usepackage{bm}
\usepackage{bbold}
\usepackage{eufrak}

\newcommand{\nn}{\nonumber}
\def \pt{\partial}

\def \L{\mathcal{L}}
\def \La{\L_{_{\!EH}}^{lin}}
\def \Lb{\widetilde{\L}_{_{EH}}^{lin}}

\def \N{\mathcal{N}}

\def \FN{F_{\!_{\N}}}
\def \and{\textmd{and}}

\begin{document}
\title{Effective models of quantum gravity induced by Planck scale modifications in the covariant quantum algebra}

\author{G.P. de Brito}\email{gpbrito@cbpf.br}
\affiliation{Centro Brasileiro de Pesquisas F\'{i}sicas (CBPF), Rua Dr. Xavier Sigaud 150, Urca, Rio de Janeiro, Brazil, CEP 22290-180}

\author{P.I.C. Caneda}\email{caneda@cbpf.br}
\affiliation{Centro Brasileiro de Pesquisas F\'{i}sicas (CBPF), Rua Dr. Xavier Sigaud 150, Urca, Rio de Janeiro, Brazil, CEP 22290-180}

\author{Y.M.P. Gomes}\email{ymuller@cbpf.br}
\affiliation{Centro Brasileiro de Pesquisas F\'{i}sicas (CBPF), Rua Dr. Xavier Sigaud 150, Urca, Rio de Janeiro, Brazil, CEP 22290-180}

\author{J.T. Guaitolini Junior}\email{jguaitolini@ifes.edu.br}
\affiliation{Centro Brasileiro de Pesquisas F\'{i}sicas (CBPF), Rua Dr. Xavier Sigaud 150, Urca, Rio de Janeiro, Brazil, CEP 22290-180}
\affiliation{Instituto Federal do Esp\'{i}rito Santo (IFES),\\ Av. Vit\'{o}ria 1729, Jucutuquara, Vit\'{o}ria, ES, Brazil, CEP 29040-780}

\author{V. Nikoofard}\email{vahid@fat.uerj.br}
\affiliation{Centro Brasileiro de Pesquisas F\'{i}sicas (CBPF), Rua Dr. Xavier Sigaud 150, Urca, Rio de Janeiro, Brazil, CEP 22290-180}
\affiliation{Departamento de Matem\'{a}tica, F\'{i}sica e Computa\c{c}\~ao, Faculdade de Tecnologia, Universidade do Estado do Rio de Janeiro,\\ Rodovia Presidente Dutra, Km 298, Polo Industrial, Resende, RJ, Brazil, CEP 27537-000}

\begin{abstract}
In this paper we introduce a modified covariant quantum algebra based in the so-called Quesne-Tkachuk algebra. By means of a deformation procedure we arrive at a class of higher derivative models of gravity. The study of the particle spectra of these models reveals an equivalence with the physical content of the well-known higher derivative gravities. The particle spectrum exhibits the presence of spurious complex ghosts and, in light of this problem, we suggest an interesting interpretation in the context of minimal length theories. Also, a discussion regarding the non-relativistic potential energy is proposed.
\end{abstract}

\maketitle


\section{Introduction} \label{sec1}

\indent 

The construction of a quantum theory for gravity consists in one of the most challenging problems of theoretical physics, since there is no experimental hints about what kind of effects we should expect from quantum corrections to the gravitational interaction. However, even without experimental evidences, a great deal of theoretical effort has been done in the last decades. In this vein, several approaches have been proposed as quantum theories of gravity, for instance, string theories, loop quantum gravity, causal dynamical triangulations, causal sets and induced quantum gravity \cite{Woodard:2009,Carlip:2001}. Nevertheless, in the present state of art, none of the preceding theories give the final word in quantum gravity.

Rather than considering a full quantum theory of gravity, one can deal with effective theories of quantum gravity, which are promising approaches to implement quantum corrections to GR \cite{Buchbinder}. However, serious problems must be faced in this approach, mainly the incompatibility between renormalizability and unitarity, which are desired features in the quantum field theory formulation. On one hand, effective theories based on GR are non-renormalizable by power counting, since the coupling constant associated with gravitation has inverse canonical mass dimension. The issue with renormalizability may be solved by introducing higher derivative terms into the action, because in such theories the UV behavior of the propagator is more convergent than in the GR case \cite{Stelle:1976gc,Modesto:2012ys,Shapiro:1997,Modesto:2012,Dona,Modesto:2014,Biswas:2012}. On the other hand, unitarity is usually lost in higher derivative theories, since it leads to unphysical massive ghosts. The riddle of the incompatibility between renormalizability and unitarity was recently explored in Ref. \cite{Accioly:2013hwa}, also relating the aforementioned features with the behavior of the non-relativistic potential energy associated with the gravitational field.

As a consequence of the association between the Planck length and the coupling constant of the gravitational interaction, the Planck scale is the natural regime in which we expect that quantum effects become most relevant to the gravitational interaction. Notwithstanding, gravity does not allow an arbitrarily amount of mass/energy in a very small region of spacetime, since it would collapse into a black hole \cite{Maggiore}. This suggests a minimal length hypothesis, \textit{i.e.} the existence of a fundamental length scale below which we cannot access. The minimal length hypothesis may be achieved by modifications in the quantum algebra of position coordinates and conjugated \textit{momenta}, which was first accomplished in a seminal paper by Snyder in 1947 \cite{Snyder:1946qz}. 

The interest on minimal length physics has increased considerably in the last two decades, specially due to certain important results within the context of string theories, loop quantum gravity and asymptotic safe gravity \cite{Seiberg:1999vs,Hossenfelder,Reuter,Percacci}. In special, we mention the non-covariant Kempf algebra \cite{Kempf:1997} which has been vastly explored in the last two decades \cite{Brau:1999,Chang:2002,Sandor:2005,Faizal:2015,Faizal:2015-IJMPA,Faizal:2016-IJMPD}. Its covariant generalization was later introduced by Quesne and Tkachuk \cite{Quesne:2006fs,Quesne:2006is}.

Remarkably the Quesne-Tkachuk algebra leads to a systematic procedure to generate fourth-derivative models, for instance: in the case spin-0 particles, minimal length corrections were implemented in the Klein-Gordon field, leading to a higher derivative theory for the scalar field \cite{Moayedi:2010vp}; for spin-1/2 fields, the deformation procedure was applied in order to construct a higher derivative version of the Dirac field \cite{Moayedi:2011ur}; some investigations were also performed in the context of electrostatic \cite{Moayedi:2012fu}, magnetostatic \cite{Moayedi:2013nba}, electrodynamics with external sources \cite{Moayedi:2013nxa} and quantum electrodynamics \cite{Silva:2016nmq}; finally, in a recent paper, Dias \textit{et.al.} explored the issue of minimal length corrections in the context of Einstein-Hilbert theory \cite{Dias:2016lkg}. Remarkably, it was recently suggested that this algebra may be viewed as an emergent effect of a supersymmetry breaking of a non-anticommutative superspace \cite{Faizal:2016}.

Realizing that there is a clear connection between minimal length deformations and fourth-derivative models, it is natural to think about this kind of deformation as a road to implement quantum corrections in the gravitational interaction. In the last few years there has been an increasing interest in modified theories of gravity including sixth or more derivative terms \cite{Modesto:2014}. In fact, this class of theories are super-renormalizable or even finite at quantum level. Nevertheless, these theories cannot be obtained by means of a deformation procedure in the context of the Quesne-Tkachuk algebra. In this paper, we propose a modification on the Quesne-Tkachuk algebra by introducing higher order corrections in the deformation parameter in order to include a larger class of higher-derivative effective theories of quantum gravity.

This paper is organized as follows: in section \ref{Quesne-Tkachuk} we introduce and explore some features of the modified Quesne-Tkachuk algebra; in section \ref{higher-derivative}, we establish a connection between the deformed version of the Einstein-Hilbert Lagrangian and the effective theories of quantum gravity; in section \ref{spectrum}, we explore some features regarding the particle spectra related with the class of effective theories obtained from the minimal length deformation of the GR Lagrangian; in section \ref{classicalprop} we study some low energy consequences of the deformed gravitational theory; finally in section \ref{Final} we present our conclusions.

Throughout this paper we use $c = \hbar = 1$, $\eta_{\mu\nu} = diag(+,-,\cdots,-)$, $R^{\mu}_{\,\,\,\nu\alpha\beta} = \pt_{\alpha}\Gamma^{\mu}_{\nu\beta} + \Gamma^{\mu}_{\alpha\lambda}\Gamma^\lambda_{\nu\beta} - (\alpha \leftrightarrow \beta)$, $R_{\mu\nu} = R^\beta_{\,\,\,\mu\nu\beta}$ and $R=g^{\mu\nu}R_{\mu\nu}$.


\section{Modified Quesne-Tkachuk algebra \label{Quesne-Tkachuk}}

\indent 

The Quesne-Tkachuk algebra is the simplest possible covariant generalization of the Heisenberg algebra that allows for a minimal length \cite{Quesne:2006fs}. We may wonder if the Quesne-Tkachuk algebra is just the first-order truncation of a more general covariant algebra. In fact, there are some proposal of higher-order algebras in the literature, for instance \cite{Pedram1,Pedram2}. A commutative spacetime regime of such algebra would also be of special interest. In order to be able to built up a new algebra, we propose the following representations
\begin{eqnarray}\label{rep}
X^{\mu}=x^{\mu}
\quad \textmd{and} \quad 
P^{\mu}=F(p^2)p^{\mu},
\end{eqnarray} 
where the lower-case position and momentum operators satisfy the usual Heisenberg algebra $[x^\mu,p^\nu] = -i\eta^{\mu\nu}$ and $[x^\mu,x^\nu] = [p^\mu,p^\nu] = 0$. For the Heisenberg algebra representations to be recovered as the low-energy limit of the \eqref{rep}, the deformation $F(p^2)$ must satisfy the condition $F(p^2) \rightarrow 1$ as $p^2 \rightarrow 0$.

From the considerations above it is possible to set up an algebra. The position and momentum commutators among themselves remain trivially null. It is necessary only to compute the position-momentum commutator
\begin{equation}\label{comm}
[X^{\mu},P^{\nu}]=-iF(p^2)\eta^{\mu\nu}-2iF'(p^2)p^{\mu}p^{\nu},
\end{equation}
where the prime denotes differentiation with respect to $p^2$. To close an algebra this must be expressed fully in terms of the new momentum operator $P^\mu$. Thus, the second contribution leads to the differential equation
\begin{equation}\label{fdiffeq}
\frac{1}{F^2}\frac{dF}{dp^2}=G(P^2).
\end{equation}
Without the form of the function $G$ we can't give an explicit solution. We study the simplest possibility where $G$ is actually a constant $G=-l^2/2$, in which case the solution to \eqref{fdiffeq} is straightforward
\begin{equation}\label{deformation}
F(p^2) = \frac{1}{1+\frac{l^2}{2}p^2}.
\end{equation}
Inverting the last equation we may express $p^2$ in terms of $P^2$ and, as a consequence, using the result in \eqref{comm} we arrive at the following algebra
\begin{subequations}\label{algebra}
\begin{eqnarray}\label{xpcomutador}
[X^\mu,P^\nu] = -i \left( \frac{1 + \sqrt{1 - 2l^2 P^2}}{2}\eta^{\mu\nu} - l^2 P^\mu P^\nu \right) ,
\end{eqnarray}
\begin{eqnarray}
[X^\mu,X^\nu] = 0 
\quad \textmd{and}\quad 
[P^\mu,P^\nu] = 0,
\end{eqnarray}
\end{subequations}
with the representations
\begin{eqnarray}\label{representacoes}
X^\mu = x^\mu \qquad \textmd{and} \qquad P^{\mu} = \frac{1}{1 + \frac{l^2}{2} p^2} p^\mu. \qquad
\end{eqnarray}

The new momentum operator representation in \eqref{representacoes}, along with the correspondence principle $p_\mu \mapsto i \pt_\mu$, leads to a deformation of the derivative operator in configuration space
\begin{eqnarray}\label{derivada_deformada}
\pt_\mu \mapsto \nabla_\mu = \frac{1}{1 - \frac{l^2}{2} \Box} \pt_\mu .
\end{eqnarray}

Notice that this deformation is ill-defined only for space-like momenta with $p^2 =- 2/l^2$. Thus no problems concerning the analyticity of the deformation procedure arises if we exclude the possibility of tachyons.

We emphasize that no approximations were made in the derivation of the algebra \eqref{algebra} and representations \eqref{representacoes}. Only a simplicity assumption $G=-l^2/2$ and the consistency condition $F(p^2) \rightarrow 1$ as $p^2 \rightarrow 0$ were needed. In this framework the algebra \eqref{algebra} is exact and holds for arbitrarily high energy. On the other hand interesting physics arises in the low energy limit, where we can power expand the r.h.s. of \eqref{xpcomutador} and equation \eqref{representacoes} in the parameter $l^2$. Each truncation of this series defines different algebras summarized in table \ref{Tabela}

\begin{widetext}
\begin{center}
	\begin{table}[h]
		\begin{tabular}{|c|c|c|c|c|}
			\hline
			~        & $[X^\mu,P^\nu]$ & $P^\mu$ & $\nabla^\mu$ & Algebra\\ \hline
			$\N = 0$ & $-i\eta^{\mu\nu}$ & $p^\mu$  & $\pt^\mu$ & Heisenberg \\ 
			$\N = 1$ & $-i\left(1 - \frac{l^2}{2}P^2\right)\eta^{\mu\nu} + il^2 P^\mu P^\nu$ & $\left(1 - \frac{l^2}{2}p^2 \right)p^\mu$ & $\left(1 + \frac{l^2}{2}\Box \right)\pt^\mu$ & Quesne-Tkachuk\\
			$\N = 2$ & $-i\left(1 - \frac{l^2}{2}P^2 -\frac{l^4}{4}P^4\right)\eta^{\mu\nu} + il^2 P^\mu P^\nu$ & $\left(1 - \frac{l^2}{2}p^2 + \frac{l^4}{4}p^2\right)p^\mu$ & $\left(1 + \frac{l^2}{2}\Box  + \frac{l^4}{4}\Box^2 \right)\pt^\mu$ & New\\ 
			$\vdots$ &$\vdots$& $\vdots$&$\vdots$ & $\vdots$ \\ 
			$\N$&$-i\left(\frac{1}{2}+\sum_{n=0}^{\N}2^{n-1}\binom{1/2}{n}(-l^2P^2)^n\right)\eta^{\mu\nu}+il^2P^{\mu}P^{\nu}$ & $\sum_{n=0}^\N \left(\frac{-l^2p^2}{2}\right)^n p^\mu$ & $\sum_{n=0}^\N\left(\frac{l^2}{2}\right)^n\!\Box^n \pt^\mu$ & New\\
			\hline
		\end{tabular}
	\caption{Nontrivial commutator and representation for a few truncations of the modified Quesne-Tkachuk algebra.}
		\label{Tabela}
	\end{table}
\end{center}
\end{widetext}

The study of truncations also suggest an origin for the construction of higher-derivative models of arbitrary order. Up until now only fourth-derivative models have been constructed from the momentum deformation operation coming from the Quesne-Tkachuk algebra, most notably among them are the Podolsky-Lee-Wick QED \cite{Moayedi:2013nxa,Silva:2016nmq} and Stelle's \cite{Dias:2016lkg} higher-derivative gravity. Second and higher order effects in these constructions (leading to sixth and higher derivatives) are more subtle due to the unavoidable appearence of spacetime non-commutativity. The novel feature of our approach is that the previous procedure can be understood as just the $\N=1$ truncation of the more general deformation coming from an also modified algebra at each order in such a way as to keep spacetime commutativity. Thus higher-derivative models of arbitrary order might be constructed from \eqref{algebra} in the same spirit as Lee-Wick and Stelle's models arise from the Quesne-Tkachuk algebra, but now with guaranteed spacetime commutativity. 

It is remarkable that this modified algebra implies a generalized uncertainty principle (GUP) which leads to a minimal length. For instance, the Quesne and Tkachuk algebra ($\N = 1$) leads to the following isotropic minimal length 
\begin{equation}\label{min_len}
(\Delta X)_{\text{min}}=l\sqrt{\frac{5}{2}\left(1-\frac{l^2}{2}\langle(P^0)^2 \rangle\right)}.
\end{equation}
Additionally, for $\N\geq2$ truncations the isotropic minimal length gets corrected by correlators of higher powers of momenta inside the square root. However, the important feature about the minimal length for $\N \geq 1$ is that it behaves as $(\Delta X)_{\text{min}} \sim l + \mathcal{O}(l^2)$, \textit{i.e.} the parameter $l$ introduced in the construction of the modified algebra is associated with the minimal length.

\section{Effective models of quantum gravity from minimal length deformed Einstein-Hilbert theory \label{higher-derivative}}
In this section we explore the possibility of constructing higher derivative model by using a deformation procedure based on the modified Quesne-Tkachuk algebra. Applying this procedure to the Einstein-Hilbert theory we arrive to a class of effective theories of quantum gravity.

Let us start with the Einstein-Hilbert Lagrangian
\begin{equation}
\L_{_{\!EH}} = \frac{2}{\kappa^2} \sqrt{|g|} \, R .
\end{equation}
Considering the usual expansion around the Minkowski spacetime, $g_{\mu\nu} = \eta_{\mu\nu} + \kappa h_{\mu\nu}$, we arrive at the following Lagrangian for the linearized version of the Einstein-Hilbert theory
\begin{align}
&\La \!=\! \frac{1}{2} (\pt_\mu h_{\nu\alpha})^2 \!-\! \frac{1}{2} (\pt_\mu h )^2 \!+\! \pt_\mu h \pt_\nu h^{\mu\nu} \!-\! \pt_\mu h^{\mu\alpha} \pt^\nu h_{\nu\alpha} \,,
\end{align}
where $h = \eta^{\mu\nu} h_{\mu\nu}$.

The deformation procedure in the derivative operator, \textit{i.e.} $\partial^\mu \mapsto \nabla^{\mu}$, leads to the deformation of the EH Lagrangian itself, $\La \mapsto \Lb$, where $\Lb$ is the minimal length deformed version for the EH theory, namely
\begin{align}
& \Lb \!=\!\! \frac{1}{2} (\nabla_\mu h_{\nu\alpha})^2 \!-\! \frac{1}{2} (\nabla_\mu h)^2\!\!+\!\! \nabla_\mu h \nabla_\nu h^{\mu\nu} \!\!-\!\! \nabla_\mu h^{\mu\alpha} \nabla^\nu h_{\nu\alpha}.
\end{align}

Now we recall that the deformed derivative operator, based in the $\N$-th order approximation of the modified algebra, is given by $\nabla_\mu = \sum_{n=0}^\N \left(\frac{l^2}{2}\right)^{\!\!n} \!\Box^n \pt_\mu$
and, therefore, the Lagrangian associated with the deformed Einstein-Hilbert theory is
\begin{align}\label{EH_deformada}
&\Lb = \La + \sum_{n=0}^{\N-1} (n+2) \left(\frac{l^2}{2}\right)^{\!\!n+1}\!\!\! \bigg( h_{\mu\nu} \Box^{n+1} \pt^\mu \pt_\alpha h^{\alpha\nu} + \nn\\
&+\frac{1}{2} h\,\Box^{n+2}h \!-\!  h\, \Box^{n+1}\pt_\mu \pt_\nu h^{\mu\nu} \!-\!\frac{1}{2} h_{\mu\nu} \Box^{n+2} h^{\mu\nu} \bigg) .
\end{align}

In fact, the above Lagrangian corresponding to the deformed version of the linearized Einstein-Hilbert theory is exactly the same as the one  obtained from the linearization of the higher derivative theory 
\begin{eqnarray}\label{label_teoria_deformada}
\mkern-30mu\tilde{\L}_{_{\!EH}} \!=\! \frac{2}{\kappa^2} \sqrt{|g|} \left[  R \!-\!  \sum_{n=0}^{\N-1} (n+2)\left(\frac{l^2}{2}\right)^{\!\!n+1} \!\!\! G_{\mu\nu} \Box^n R^{\mu\nu} \right] \!\!,
\end{eqnarray}
where $G_{\mu\nu} = R_{\mu\nu} - \frac{1}{2}g_{\mu\nu}R$ is the Einstein tensor.


\section{Particle spectrum and tree-level unitarity\label{spectrum}}
Now let us analyze the particle spectrum associated with the deformed Lagrangian \eqref{label_teoria_deformada}. 
In order to investigate the tree-level unitarity, we compute the saturated propagator \cite{Accioly:2013hwa}, namely
\begin{align}\label{saturated_propagator}
\mkern-15mu SP(k) \!=\! \left( \frac{1}{k^2} \!+\! \frac{l^2}{2} \frac{\FN(k^2;1;2)}{\FN(k^2;0;1)} \right) \left(T_{\mu\nu}T^{\mu\nu} \!-\! \frac{1}{2}T^2 \right),
\end{align}
where we use the following definition
\begin{eqnarray}
\FN(k^2;\zeta;\xi) = \sum_{n=0}^{\N-\zeta} (n+\xi) \left( \frac{l^2}{2} \right)^{n} (-k^2)^n.
\end{eqnarray}
Since we are interested in the particle spectrum we have to look at the pole structure of the saturated propagator, in this vein it is not difficult to see that there is a simple pole at $k^2=0$ (for all $\N$) and that there are additional poles identified as the zeros of $\FN(k^2;0;1)$. 

Let us first discuss the pole at $k^2 = 0$, this pole corresponds to a massless particle with spin-2, to be identified with the usual graviton. To verify the tree-level unitarity, we compute the residue of the saturated propagator at $k^2 = 0$, leading to the following result \cite{Accioly:2002}
\begin{align}
Res(SP)|_{k^2=0}  = \left(T_{\mu\nu}T^{\mu\nu} - \frac{1}{2}T^2 \right)\bigg|_{k^2=0} >0,
\end{align}
hence, the massless particle associated with the pole $k^2 = 0$ is ``healthy'' in the sense of tree-level unitarity.

\begin{figure}[h!]
\centering 
\qquad\quad\includegraphics[width=8cm]{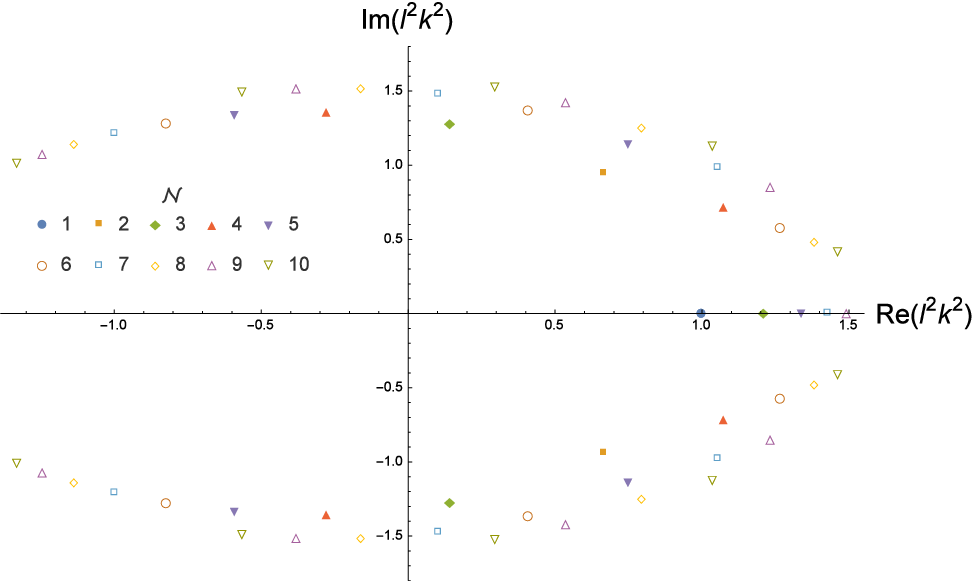}
\caption{\label{fig:i} Structure of non-trivial poles of the propagator on the complex plane.}
\end{figure}

The zero structure of the function $\FN(k^2;0;1)$ cannot be found for an arbitrary $\N$, however one can use numerical techniques in order to determine the zero associated with an specific $\N$. In figure \ref{fig:i} we plot the zero structure of $\FN(k^2;0;1)$ for values of $\N$ ranging from $\N=1$ up to $\N = 10$. As one can see the zeros are not restricted to the real axis, they are distributed on the complex plane.

In fact, it can be demonstrated that the zero structure of $\FN(k^2;0;1)$ has the following distribution on the complex plane:
\begin{itemize}
\item[i)] $\N = \textmd{Odd} \,\Rightarrow $  A single root at $k^2 = m^2$, ($m^2 >0$), and $\N-1$ complex roots (``complex masses'');
\item[ii)] $\N = \textmd{Even} \,\Rightarrow $  $\N$ complex roots (``complex masses'').
\end{itemize}
As one can see in the above results, there are two distinct kinds of poles in the propagator, the first being real poles (occurring only for $\N$ odd) and the second being complex poles (occurring for $\N$ both odd and even).
In the case of real poles, the same analysis done for the pole $k^2 = 0$ may be performed and it can be verified that it corresponds to a massive ghost particle. It is important to note that this result is already expected when we deal with higher derivative models.

Now let us discuss the complex poles. Although it is not an usual situation, we intend to argue that it could lead to a very interesting physical scenario. The usual interpretation relates the poles of the propagator with the masses of the physical particles described by the theory. However, for complex poles this identification cannot be made, since masses must be observables. A possible interpretation may be to identify the imaginary part of the poles with the decay rate of unstable particles \cite{Kuksa:2015,Turcati:2016}. In this vein we refer to a sequence of papers by H. Yamamoto \cite{Yamamoto:1969,Yamamoto:1970,Yamamoto:1976}, where a careful analysis of the physical relevance of propagators with complex \textit{}poles was performed. It was argued that the inclusion of complex poles in the propagator may lead to problems in the causal structure of the theory, more precisely, \textit{microcausality}. By \textit{microcausality} we mean \textit{causality} for an arbitrarily infinitesimal time-like interval. Nonetheless, it was noted that \textit{microcausality} is not an indispensable concept from the observational point of view, since we cannot access arbitrarily small spacetime intervals, in this sense only \textit{macrocausality} is required. Indeed the causal structure was studied by Yamamoto for a theory described by a scalar field containing complex poles and in this case the \textit{macrocausal} structure is respected, although \textit{microcausality} is violated. The same situation arises in the theory \eqref{label_teoria_deformada}, though the region where \textit{microcausality} violation takes place is related to the minimal length.

A natural question which arises is how to identify the boundary between \textit{micro} and \textit{macrocausality}. We shall argue that the generalized uncertainty principle establishes a natural boundary between both concepts. As it was mentioned before, the complex poles in the propagator could jeopardize the causal structure of the theory due to the presence of spurious particles with complex ``masses'' in the spectrum. However, since the imaginary part of the complex poles might be associated with the decay rates of these particles, there will be a finite region where they can propagate, since it corresponds to unstable particles. In fact, it can be (numerically) verified that the imaginary part of the complex poles of the propagator \eqref{saturated_propagator} are of order $\sim 1/l^2$ and, as a consequence, the life-time of these modes are of the order $\tau \sim l$. Additionally, it can be shown that the group velocity respects the condition $v_g < 1$. Therefore, the decay length $\Gamma $ of these particles also are of order $\sim l$. 

\begin{figure}[h!]
\centering 
\includegraphics[width=6cm]{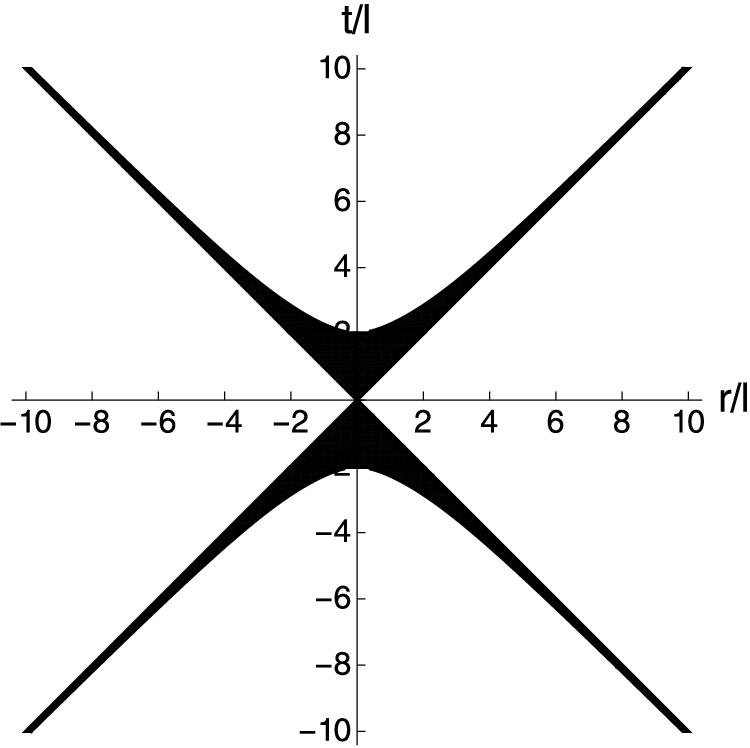}
\caption{\label{fig:ii} Space-time region where the microcausality violation occurs (in the case $\N = 2$).}
\end{figure}

Although there is a specific region in the light-cone where these modes can survive, see figure \ref{fig:ii}, if we deal with the concept of minimal length the problems with \textit{microcausality} are avoided, since the infinitesimal time-like interval where \textit{causality} is violated is located inside a region with characteristic scale of the order $ \sim l$. Considering that the relation \eqref{min_len} introduces a minimum observable length, \textit{i.e.} a fundamental uncertainty in the position measurements, the undesired consequences of the complex modes would not be detectable. Finally, regarding the contribution of these modes to the $S$-matrix, we refer to the papers by Modesto and Shapiro \cite{Modesto:2016,Shapiro:2016}, where the authors claimed that the complex poles do not contribute to the (non-)unitarity of the theory, since they correspond to unstable particles. In this case, the theory may be viewed as unitary in the Lee-Wick sense. 

\section{The non-relativistic potential energy \label{classicalprop}}

Since there is no experimental hints on what we should expect from a theory of quantum gravity, we usually look for classical corrections coming from the quantum regime. An interesting probe is the non-relativistic potential energy, which brings some new contribution to the usual Newtonian potential energy. In order to compute this quantity we follow reference \cite{Accioly:2015}, where the authors derived a straightforward expression to compute the non-relativistic gravitational potential energy associated with two particles, with respective masses $M_1$ and $M_2$, separated by a distance $r \equiv |\vec{r}\,|$. Without going further with explicit calculations we arrive at the following result for $\N$ even
\begin{subequations}
	\begin{eqnarray}
	E(r) = - \frac{GM_1 M_2}{r} \bigg( 1 + 2\sum_{l=1}^{\N / 2} g_l(r) \bigg) ,
	\end{eqnarray}
	and, for $\N$ odd
	\begin{eqnarray}
	\mkern-15mu E(r) = - \frac{GM_1 M_2}{r} \bigg( 1 + 2\!\!\!\!\!\sum_{l=1}^{(\N-1) / 2}\!\!\! g_l(r) + c_0 e^{-m r} \bigg) \!,
	\end{eqnarray}
\end{subequations}
where $g_l(r) = e^{-\alpha_l r} \Big( a_l \cos (\beta_l r) + b_l \sin(\beta_l r)\Big)$. Also, it is important to stress out that all the constants $a_l$, $b_l$, $c_0$, $\alpha_l$ and $\beta_l$ are determined in terms of the minimal length parameter.
\begin{figure}[h!]
\centering 
\includegraphics[width=8cm]{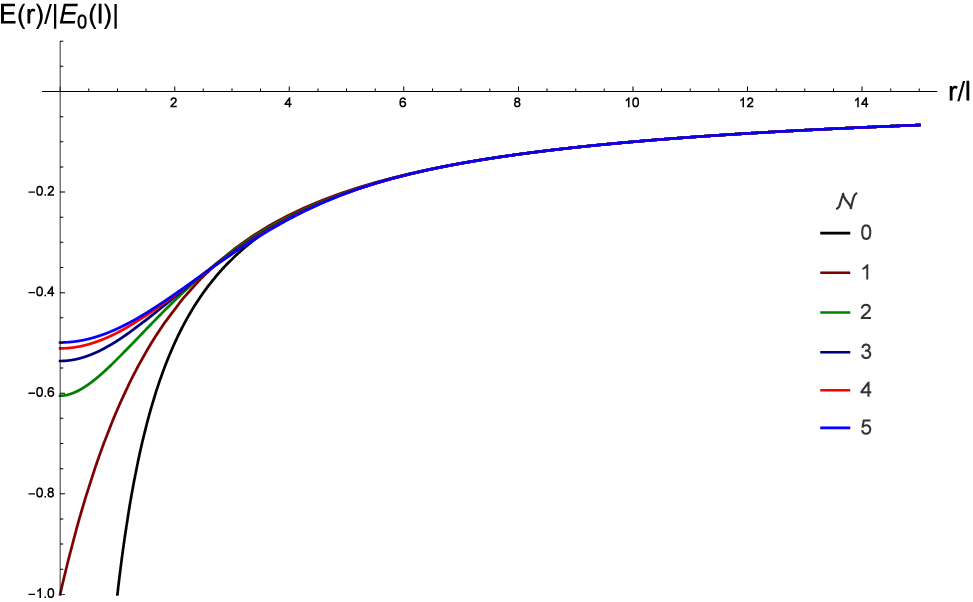}
\caption{\label{fig:iii} The non-relativistic gravitational potential energy for different values of $\N$. We have defined $E_0(l) = - G M_1 M_2/l$.}
\end{figure}

\begin{figure}[h!]
\centering 
\includegraphics[width=8cm]{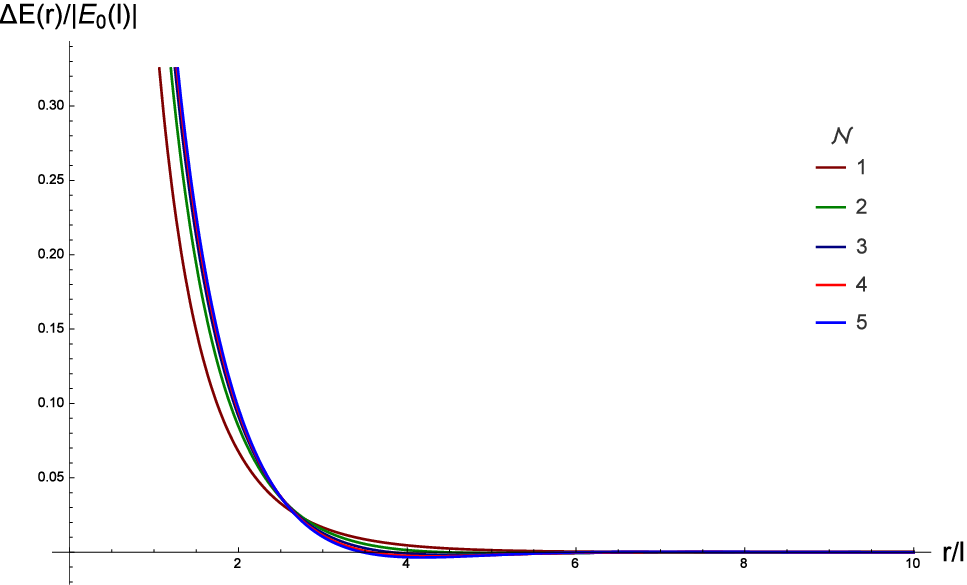}
\caption{\label{fig:iv} The contribution of the extra terms to the non-relativistic gravitational potential energy for different values of $\N$. We have defined $E_0(l) = - G M_1 M_2/l$.}
\end{figure}

As one can see in figure \ref{fig:iii}, the non-relativistic potential energy (for $\N \neq 0$) has a remarkable behavior at $r=0$, it converges to a finite value. The mechanism behind the cancellation of Newtonian singularity was recently explored in references \cite{Tiberio:2015,Breno:2016}. The idea is quite simple, in fact the non-relativistic potential energy may be split into two contributions, namely $E(r) = E_0(r) + \Delta E(r)$, where $E_0(r) = - G M_1 M_2/r$ stands for the usual Newtonian potential energy associated with the physical graviton and $\Delta E(r)$ represents the remaining contribution coming from the non-trivial poles. Indeed, in figure \ref{fig:iv} we have plotted the contribution $\Delta E(r)$ and, as one can see, it has a repulsive behavior when we approach $r=0$, canceling the Newtonian singularity at the origin. It's worth to mention that this repulsive behavior is valid for arbitrary distances in the case $\N = 1$, while in the case $\N \geq 2$ there is a small region where the potential energy becomes negative due to the presence of oscillating terms. 

\begin{figure}[h!]
\centering 
\includegraphics[width=8cm]{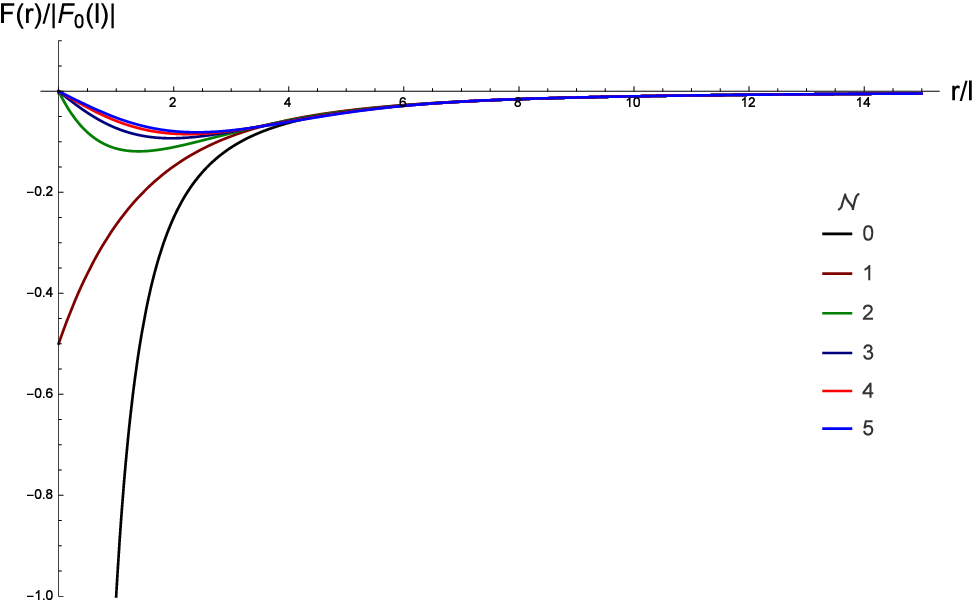}
\caption{\label{fig:v} The gravitational force for different values of $\N$. We have defined $F_0(l) = - G M_1 M_2/l^2$.}
\end{figure}

Another interesting feature of the phenomenological implications of the gravitational interaction at small distances may be seen when we look for the Newtonian-like gravitational force. In fact, the radial component of the force can be directly computed from the non-relativistic potential energy by means of the usual formula $F(r) = - \pt E(r)/ \pt r$. In figure \ref{fig:v} we have plotted the gravitational force for diverse values of $\N$. As one can see, the behavior at the origin is surprising for the cases $\N \geq 2$: the gravitational force is null at $r=0$. From the phenomenological point of view, this is not a problem, since we cannot probe the gravitational interaction for small distances. Indeed, we call the attention to reference \cite{Narain:2012}, where the author investigated the short distance freedom of quantum gravity by means of a renormalization group analysis. Nonetheless, the gravitational force reproduces the Newtonian $1/r^2$ behavior for large distances compared with the minimal length.

\section{Final remarks \label{Final}}

Motivated by a series of papers by Moayedi \textit{et. al.} \cite{Moayedi:2010vp,Moayedi:2011ur,Moayedi:2012fu,Moayedi:2013nba,Moayedi:2013nxa}, Silva \textit{et.al.} \cite{Silva:2016nmq} and Dias \textit{et.al.} \cite{Dias:2016lkg}, we apply the modified Quesne-Tkachuk algebra to the construction of higher-derivative models by means of a deformation procedure. As it was demonstrated in section \ref{higher-derivative}, the deformation of the Einstein-Hilbert theory gives a class of theories containing the same particle content of the higher-derivative gravities constructed by the inclusion of $R\,\Box^n R$ and $R_{\mu\nu}\Box^n R^{\mu\nu}$ terms in the Lagrangian. 

One of the most important problems of higher-derivative models is the presence of spurious states with negative norm, called ghosts, which contribute to the non-unitarity of the $S$-matrix. The analysis of the particle spectra associated with the deformed models appears to give a promising result. In the case of $\N$ even, one can evade the problem of tree-unitarity by a careful interpretation of complex poles in the propagator, combined with the interpretation of a minimal uncertainty in the position measurement. In this vein, we follow the interpretation of Yamamoto  for the complex poles \cite{Yamamoto:1969,Yamamoto:1970,Yamamoto:1976} which compromises the \textit{causal} structure of the theory, but only in a microscopic region where we cannot access by experimental means. This interpretation may be supported by the minimal length hypothesis, which determines a physical limit in the measurement of distances. Notwithstanding, the \textit{causal} structure for large distances, compared with the minimal length, is maintained. In this case the theory is unitary in the Lee-Wick sense, since all the massive ghost-like states correspond to unstable particles \cite{Modesto:2016,Shapiro:2016}. The same discussion regarding the complex poles applies to the case of $\N$ odd, however, in this case the problem of non-unitarity should be faced, since there is a massive ghost (associated with a real pole) in the spectrum.

In order to investigate the low energy consistency of the deformed theory we have computed, in Sec. \ref{classicalprop}, the non-relativistic gravitational potential energy, inspired by the paper of Ref. \cite{Accioly:2015}. As a result we found that the Newtonian potential energy is extended by some terms associated with the additional poles of the propagator, the real poles contributing with Yukawa-like terms and the complex ones contributing with damped-oscillating portions. Remarkably, except for the case $\N=0$, the non-relativistic potential energy is finite at $r=0$ \cite{Tiberio:2015,Breno:2016}.

Remarkably, higher-derivative models of gravity have received considerably attention over the last decades, mainly due to their renormalizability properties. Since the seminal work by Stelle \cite{Stelle:1976gc}, it has been known that fourth-derivative models of gravity are renormalizable. In this spirit, a preliminary analysis of the primary UV divergences of the Feynman diagrams indicates a possible (super-)renormalizability of the deformed Einstein-Hilbert theory. However, a more careful analysis is underway in order to investigate if the specific couplings of the Lagrangian \eqref{label_teoria_deformada} are preserved after addition of counter-terms.

Natural questions that arise within the present work regards the fate of the real massive ghost found only on odd truncations and, also as a consequence thereof, the behavior of the gravitational potential energy at scales close to the minimal length $r/l\sim 1$. To progress towards the answers we must study the high energy limit of the model by applying the full derivative deformation \eqref{derivada_deformada} to the Einstein-Hilbert theory. In this limit we must fully face the non-local aspect of a minimal length theory, which in the proposed algebra is signaled by the presence of the d'Alembertian in the denominator. It would also be interesting to investigate how the proposed algebra and deformed theory affects recent results on higher-derivative black holes \cite{Faizal:2015-IJMPA,Majumder:2011,Bambi:2017}. 

Remarkably, along this paper we restricted ourselves to investigate the consequences of the truncated versions of the proposed modified algebra. The complete representation that we found in section \ref{Quesne-Tkachuk} results in the appearance of non-local terms. In such a case, we can also demonstrate that the deformed theory corresponds to the linearized version of a non-local theory of quantum gravity. Non-local theories received considerable attention in the last few years as being an interesting framework to reconcile renormalizability and unitarity in quantum gravity (see reference \cite{Modesto:2017} for an updated review). In a future publication we intend to investigate the properties of the non-local theory obtained with the complete representation of the modified Quesne-Tkachuk algebra.

\section*{Acknowledgments}

The  authors  are  grateful  to  J.A.  Helay\"{e}l-Neto  for  interesting  discussions and for reading the manuscript.  This work was funded by the Brazilian National Council for Scientific and Technological Development (CNPq). 


\end{document}